\documentclass[review]{elsarticle}

\usepackage{multirow}
\usepackage{amsmath}
\usepackage{breqn}
\usepackage{amssymb}
\usepackage{lineno,hyperref}
\modulolinenumbers[5]
\usepackage{lscape}
\usepackage{longtable}
\usepackage{color}
\usepackage{graphicx}

\usepackage[textwidth=340pt,textheight=550pt]{geometry}

\journal{arXiv}

\begin{document}
	
	\begin{frontmatter}
		
		\title{Natural Time-Series Analysis and Vedic Hindu Calendar System}
		
		
		\author[]{Neeraj Dhanraj Bokde\corref{mycorrespondingauthor}}
		\cortext[mycorrespondingauthor]{Corresponding author}
		\ead{neerajdhanraj@cae.au.dk}
			
		\address{Department of Civil and Architectural Engineering, Aarhus University, 8000, Denmark}
		
		\begin{abstract}
			
Worldwide, calendars are classified into three categories, solar, lunar, and lunisolar, based on motions of Sun, Moon, and both, respectively. Being lunisolar, the Vedic Hindu calendars are capable of considering both solar and lunar activities. Therefore, these calendars can have potentially an upper hand over solar and lunar calendars. Several natural activities on the earth are due to the influences of the Sun and Moon, and the Hindu calendars are able to observe these patterns. Time series analysis plays a very crucial role in day-to-day applications and is one of the major components in the field of data science. The modern computations (including time series analysis) are performed with the Gregorian (solar) calendars. In this paper, the potential of the Hindu calendar is discussed from the time series analysis point of view. Some logical and experimental comments are discussed in this paper to motivate to dig further possibilities in the domain.
		\end{abstract}
		
		
	\end{frontmatter}
	
	
\section{Introduction}
	Time series analysis plays an important role in day-to-day life in several applications, such as economics, health, finance, genetics, energy, environment, and many more. Time-series forecasting is one of the major components in the field of data science, because of its nature and applications. The time series forecasting methods need to discover the patterns within the time series without any reference parameters or values, and this makes the task challenging, unsatisfactory results, and unachievable targets, sometimes. After the new age advancements in computational technologies, researchers are developing new technologies, methodologies, frameworks, testbenches, and algorithms to ease down the procedures, minimizing efforts, reducing the execution time, and achieve magnificent accuracies in the forecast results \cite{phillips2020advances}. A new age forecasting method is been proposed recently, which is excellent in finding significant patterns within the time series \cite{RJ-2017-021} (for example, ARIMA, PSF, RNN, etc.).
	
	The time-series analysis processes such as cleaning, imputation, or forecasting are the procedures that demand the historic time series and are evaluated in several ways. In common words, a time series is a data type based on time, as a sequence of discrete-time data. It is a sequence taken at successive equally spaced points in the time domain. The majority of the time series analyzing methods are of a similar structure. These methods dig the patterns within the time series and try to understand the alignment of the historic nature of the time series to search the patterns within. The modern computations are performed in the time scale provided by the Gregorian calendar. The Gregorian calendar is based on the relative behavior of the Sun and Earth, hence known as a solar calendar. In this calendar, a year is measured as a time taken by the earth to complete an orbit around the Sun, which is documented to be precisely 365 days, 5 hours, 48 minutes, and 46 seconds \cite{meeus1992history}. The popular Gregorian calendar consists of 12 months with 30 or 31 days each month (except February, which is of 28 or 29 days). The number of days in February is assigned to 29 after every four years (with some exceptions) to manipulate the difference between the actual length of a year and the length of the year provided by the Gregorian calendar \cite{richards1999mapping}.
	
	It is worth noticing that there are rules defined for the number of days in each month of a year in the Gregorian calendar, but a justified logic behind it is not available. Whereas, the Vedic Hindu calendars are based on the motion of both Sun and Moon. These calendars are solar year long, which is the time taken by earth to revolve around the Sun as well as the months are lunar, which is the time taken by the Moon around the earth. These lunar months and solar years are well balanced in the Hindu calendar. The difference between them is scientifically managed by introducing an additional month after every 30 months (known as Adhikmaas), maintaining the same length of each month, that is 30 days. 
	
\section{Hindu calendar for time-series analysis}
	There are several different phenomena (such as 12 Constellations, 12 Zodiac signs, and many others) are assigned with the Hindu calendar. Many of these points prove that the Hindu calendars are equally scientific as that of the Gregorian one and the question arises, what if we give a try on the Hindu (lunisolar) calendar to handle the time factor in time series analysis. There are several logical and experimental reasons, which are commented for this question as follows:
	
	\begin{enumerate}
		\item Several natural phenomenons such as lunar rhythms which are responsible for several activities in birds and animal's life (such as, reunion, migrations, mating periods, etc), tidal effects on seas, and major climate fluctuations \cite{dorminey2009without}. Besides, some studies have claimed that the could not be possible on earth in the absence of the Moon \cite{dorminey2009without}. Similarly, Sun is responsible for numerous climatic and natural changes on earth, for example, solar wind, temperature, weather, seasons, eventually the wind, rain, and several other natural behaviors on the earth \cite{rampino1984terrestrial}. It proves that both Sun and Moon have a great influence on the behavioral activities on earth, and when we attempt to document them in numeric formats, we simply form a time series. The time-series format in the Gregorian (solar) calendar can relate the patterns of the year, but what if we try to format the natural time series with the Hindu (lunisolar) calendar format? Can we extract both lunar (Moon) and solar (Sun) based patterns within the time series more efficiently with the Hindu (lunisolar) calendars?
		
		\item Computers make analysis easy and fast, but they do not have their sense. When a time series is provided, the computer handles it with the Gregorian calendar structure. It automatically formats the time series in the unequal length of months (that is 30 or 31), which leads to uneven pattern sequence formations. There are several time-series analysis applications, which demand equal divisions of the whole year time series in equal intervals such as monthly Full Load Hour (FLH) scheduling techniques in \cite{bokde2020graphical}. It is not possible to segment the time series in exactly equal parts, hence an assumption need to be used that each month is of 30 or 31 days.
		Similarly, it is expected to find the similarity or correlation coefficient between two consecutive months, the unequal length of months usually be an obstacle. Therefore, several assumptions are always used in most of these month-wise time series analysis.
		So, what if we use a Hindu (lunisolar) calendar, which divides a year into 12 months of equal lengths and adds an additional month of the same length to compensate for the difference between lunar and solar periods? Can we segment the time series in equal parts with more possibilities of synchronous patterns?
		
		\item Days and months in the Hindu calendar are based on Moon activities. The length of lunar days (known as tithis) is the time taken by the Moon to increase its longitude by 12$^0$ over the earth's longitude. Therefore, the length of a day in the Hindu calendar is not fixed as it is in the Gregorian one. The length of these tithis (days) varies from 20 to 27. We have never exercised to find the patterns and their relations based on lunar days (tithis) segmentation.
		Can we observe more realistic, accurate, or synchronous patterns with the time series segmentation with the lunar days (tithis)? 
		
		\item So far, we do not have a perfect answer to the above questions and yet, there is no such tool to perform time series analysis with the Hindu calendar, but a simple analysis and comparison discussed in this section can motivate us to dig this possibility further.
		\begin{figure}
			\includegraphics[width=12.5cm]{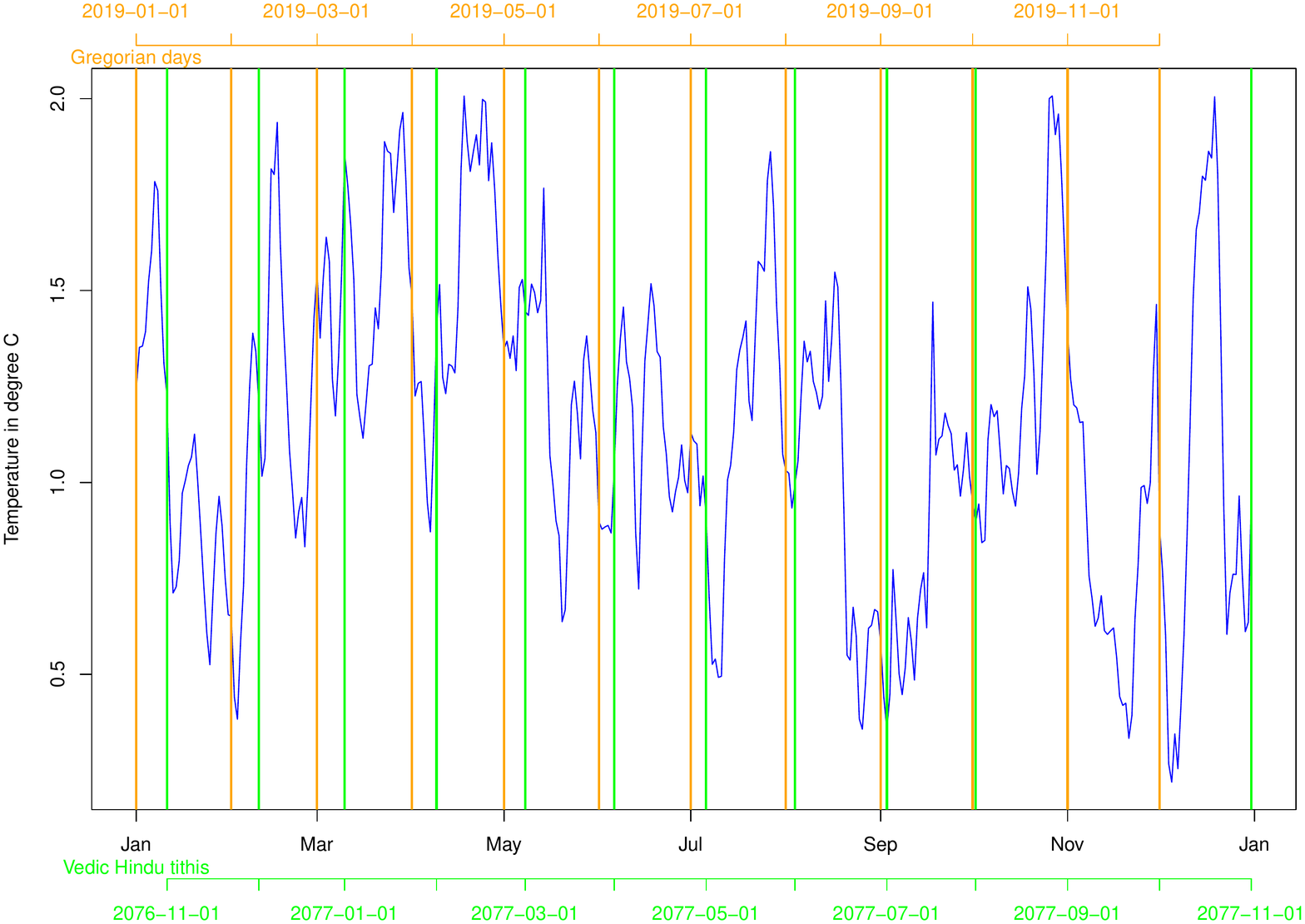}
			\caption{Segmentation of a year-long time-series into Gregorian and Hindu calendar months.}
			\label{fig:1}       
		\end{figure}
		
		\begin{figure}
			\includegraphics[width=10cm]{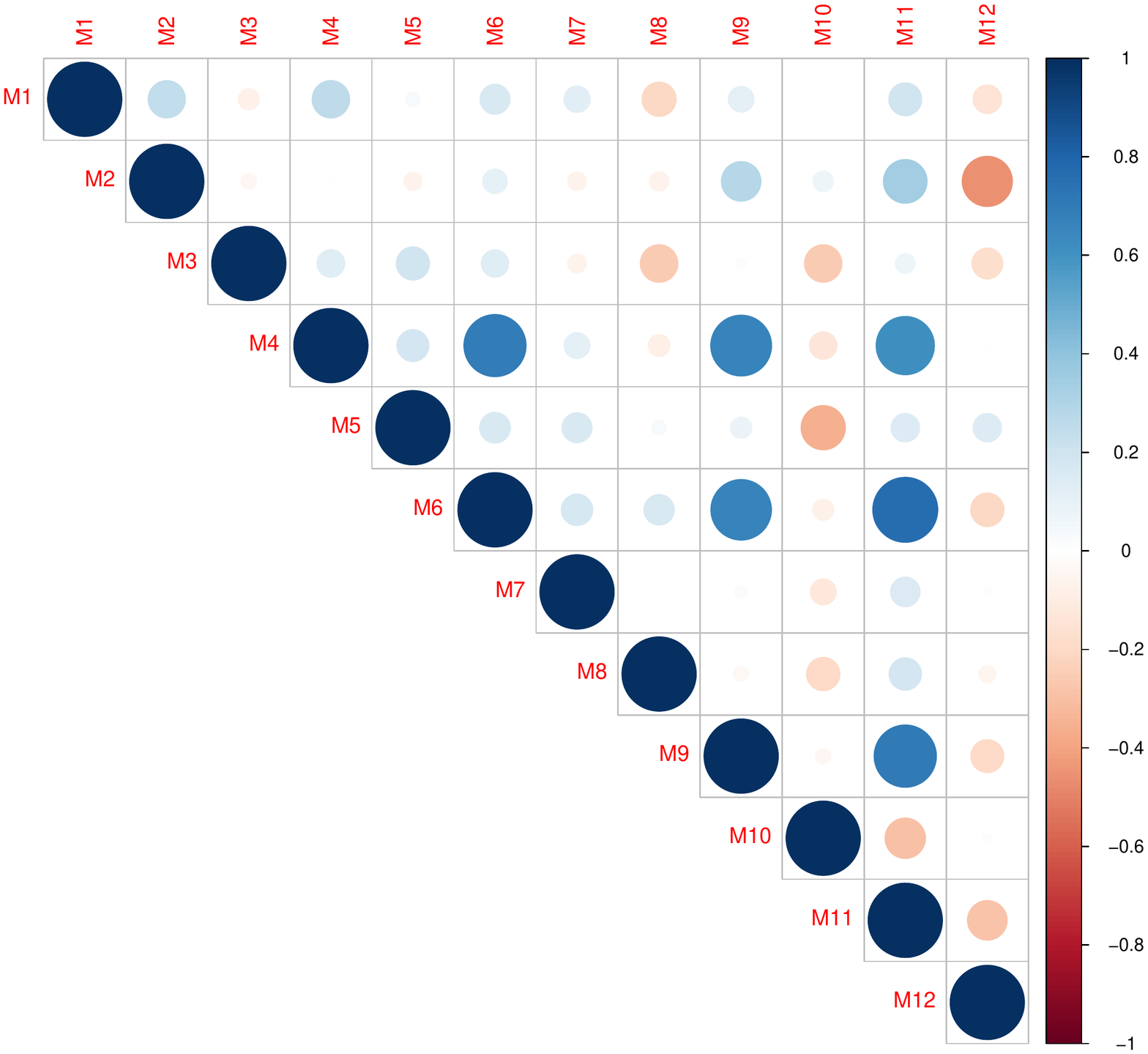}
			\caption{Correlation coefficient between segmented Gregorian months.}
			\label{fig:2}       
		\end{figure}
		
		\begin{figure}
			\includegraphics[width=10cm]{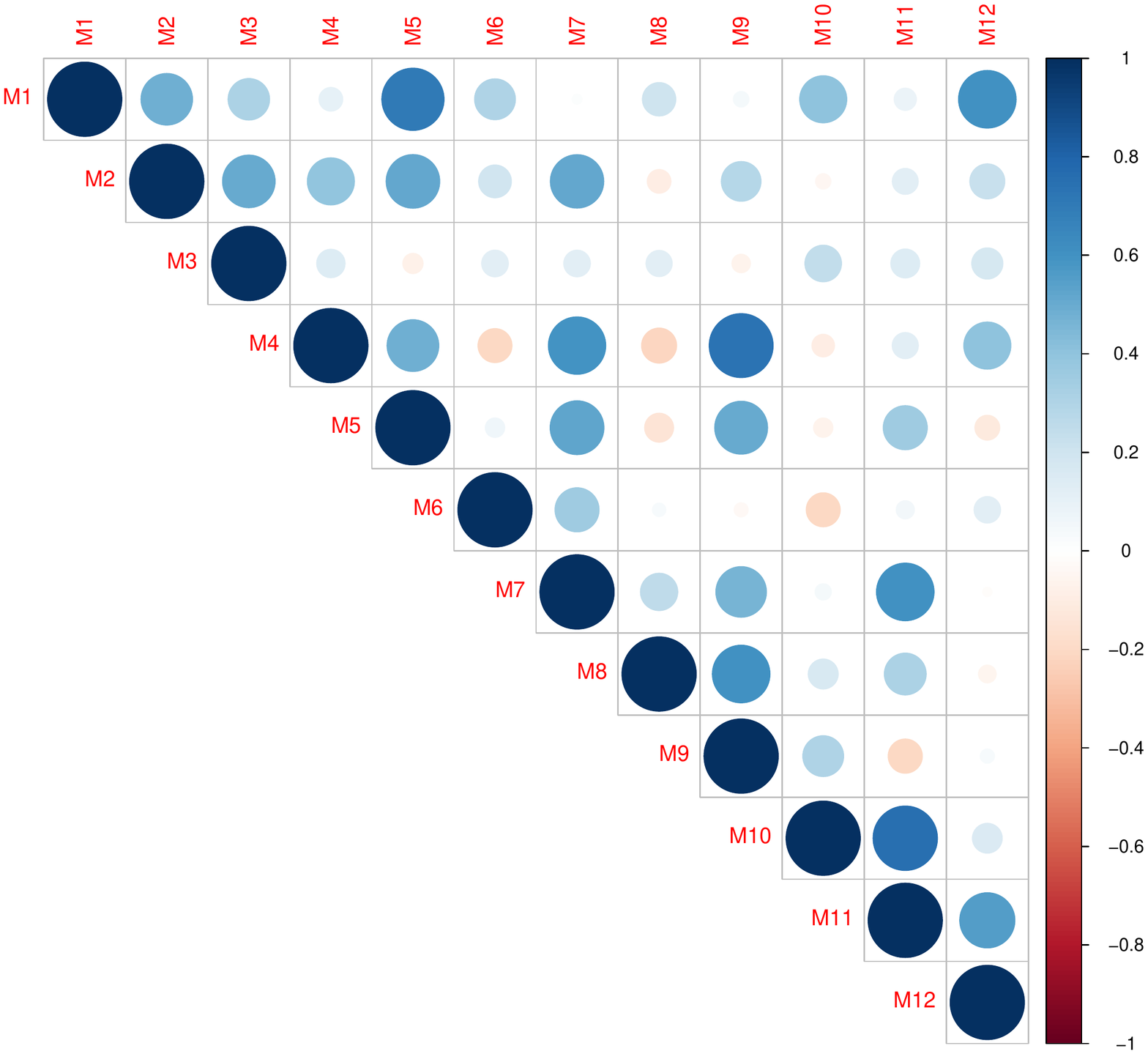}
			\caption{Correlation coefficient between segmented Hindu calendar months.}
			\label{fig:3}       
		\end{figure}
		
		Figure \ref{fig:1} represents a sample daily temperature (in $^0$C) time series for a year, and it is segmented into 12 months based on Gregorian (in orange) and Hindu-Saka system (in green) calendar. The correlation coefficient between these 12 segmented months using Gregorian and Hindu calendars are shown in Figures \ref{fig:2} and \ref{fig:3}. The darker and bigger dots are more favorable and representing a stronger relationship between the respective two months. It can be seen that Hindu calendar based segmentation (Figure \ref{fig:3}) has more number of such blue dots. This can be valuable proof that the Hindu calendar may improve the time-series analysis accuracies, but not the sufficient or enough one. More efforts in developing new tools, techniques, data types and detailed studies are desired to look deep into such possibilities.
	\end{enumerate}
	
\section{Future perspectives}
	This paper is proposed to understand the possibilities in updating or replacing the existing time measuring units in time series analysis (and data science) tools. Following are some of the future research perspectives in this domain:
	
	\begin{enumerate}
		\item Development of new data types for Date (for example, R and Python packages, which can transform existing Gregorian dates into Hindu (lunisolar) dates).
		\item Development of new time series analysis techniques that can be more suitable for Hindu calendar date patterns.
		\item Proposing new techniques that can work simultaneously on Gregorian and Hindu calendar systems.
		\item Exploring possibilities of new techniques that can extract solar and lunar patterns from a natural time series using lunisolar Hindu calendars.
	\end{enumerate}

	
	\bibliography{ref}

\begin{thebibliography}{1}
\expandafter\ifx\csname url\endcsname\relax
  \def\url#1{\texttt{#1}}\fi
\expandafter\ifx\csname urlprefix\endcsname\relax\def\urlprefix{URL }\fi
\expandafter\ifx\csname href\endcsname\relax
  \def\href#1#2{#2} \def\path#1{#1}\fi

\bibitem{phillips2020advances}
G.~Phillips-Wren, A.~Esposito, L.~C. Jain, Advances in data science:
  Methodologies and applications (2020).

\bibitem{RJ-2017-021}
N.~Bokde, G.~Asencio-Cortés, F.~Martínez-Álvarez, K.~Kulat,
  \href{https://doi.org/10.32614/RJ-2017-021}{{PSF: Introduction to R Package
  for Pattern Sequence Based Forecasting Algorithm}}, {The R Journal} 9~(1)
  (2017) 324--333.
\newblock \href {https://doi.org/10.32614/RJ-2017-021}
  {\path{doi:10.32614/RJ-2017-021}}.
\newline\urlprefix\url{https://doi.org/10.32614/RJ-2017-021}

\bibitem{meeus1992history}
J.~Meeus, D.~Savoie, The history of the tropical year, Journal of the British
  Astronomical Association 102~(1) (1992) 40--42.

\bibitem{richards1999mapping}
E.~G. Richards, Mapping time. The calendar and its history., 1999.

\bibitem{dorminey2009without}
B.~Dorminey, Without the moon, would there be life on earth, Scientific
  American 21 (2009).

\bibitem{rampino1984terrestrial}
M.~R. Rampino, R.~B. Stothers, Terrestrial mass extinctions, cometary impacts
  and the sun's motion perpendicular to the galactic plane, Nature 308~(5961)
  (1984) 709--712.

\bibitem{bokde2020graphical}
N.~Bokde, B.~Tranberg, G.~B. Andresen, A graphical approach to carbon-efficient
  spot market scheduling for power-to-x applications, Energy Conversion and
  Management 224 (2020) 113461.

\end{thebibliography}

\end{document}